# [1]Multi-objective control strategy of Electro-Mechanical Transmission Based on Driving Pattern Division

Yanbo Li[1], Jinsong Li[2], Zongjue Liu[2], Riming Xu[3*]

1. Johns Hopkins University, Department of Civil and System Engineering, Baltimore, MD, US

2. Mechanical and Electrical Engineering College, Hainan University, Haikou, CN

3. King Abdullah University of Science and Technology, Physical Sciences and Engineering Division, Thuwal, Makkah, Province, SA

**Abstract:** Based on the driving requirement and power balance of heavy-duty vehicle equipped with Electro-Mechanical Transmission (EMT), optimization goals under different driving patterns are put forward. The optimization objectives are changed into a comprehensive optimization target based on the method of weighting, which is calculated by using analytic hierarchy process (AHP) under different working conditions. According to theory of Dynamic Programming (DP), a multi-object control strategy of DP under different driving patterns is proposed. This strategy is verified by simulation and contrasted with rule strategy, the results show that comprehensive performance is significantly enhanced, and the fuel economy is highly improved especially.

[1] Corresponding Author

Riming Xu

King Abdullah University of Science and Technology (KAUST), Thuwal 23955-6900, Saudi Arabia

Email: riming.xu@kaust.edu.sa

**Nomenclature**

*Abbreviation*

EMT             Electro-Mechanical Transmission

AHP             Analytic Hierarchy Process

DP              Dynamic Programming

# 1. Introduction

With the development of electro-mechanical transmission(EMT) technology and Electric transmission technology, the EMT tracked vehicle has become a research hotspot in various countries. It can fulfill electric and mechanical transmission and makes full use of advantages of power sources such as engine or power battery pack. The EMT vehicles have great potential of application and future development. The controllers of EMT vehicles mostly operate on either a rule-based or an optimization-based algorithm, each having its own advantages and disadvantages[1].

The control strategy is the technical core and design difficulty of EMT vehicle research.Control system serves to coordinate power flow from source to load. An optimized control system will improve vehicle efficiency, stability and performance.[2] Under the premise of meeting vehicle's power requirement and other basic performance requirements, according to driving conditions and characteristics of powertrain components, by using the energy-saving mechanism of EMT to give a full play to the energy-saving potential of the system and to achieve the objectives of fuel economy, emission, etc., is a typical multi-objective control issue.

The global optimal control strategy[3-5] is currently the focus of research and development in the field of hybrid vehicle energy management. However it faces two problems. One is the current optimal control algorithm can only be applied to global optimal control problems with known driving conditions. It cannot be applied to real time control yet. The second is, for multi- objective optimization problems, in order to achieve optimal objectives, multiple objectives are simplified to a single objective in most studies. The basic idea is:

the relative importance of each objective is analyzed on different conditions, and the most crucial objective of the current condition is selected to be the optimal objective. Other objectives are regarded as constraints. This method is actually still a single-objective optimal control method and the multi-objective optimal control strategy is not obtained. In this paper, a comprehensive optimization objective of different conditions is set up by using the analytic hierarchy process(AHP), and the optimal control strategy is realized on the basis of dynamic programming(DP).

In Ref. [6] the design procedure starts by defining a cost function, such as minimizing a combination of fuel consumption and selected emission species over a driving cycle. Dynamic programming is then utilized to find the optimal control actions including the gear-shifting sequence and the power split between the engine and motor while subject to a battery SOC-sustaining constraint. In Ref. [7], two control algorithms are introduced: one based on the stochastic dynamic programming method, and the other based on the equivalent consumption minimization strategy. Both approaches determine the engine power based on the overall vehicle efficiency and apply the electrical machines to optimize the engine operation. The performance of these two algorithms is assessed by comparing against the dynamic programming results, which are non-causal but provide theoretical benchmarks for other implementable control algorithms. In Ref. [8], The whole vehicle system may be abstracted to one consisting of two energy sources, one of them rechargeable and the other consumable, that feed or receive energy from an energy consumer. The authors derive the power split between the two sources such that fuel consumption is minimized, while the vehicle performs a given velocity cycle. Bounds on the power flows from both sources are considered. The problem is posed as a finite horizon dynamical optimization problem with constraints and solved by a dynamic programming approach.

Many researchers combines dynamic programming and other methods of control strategies, and some researchers improve the algorithm based on the original dynamic programming. In Ref. [9], The online intelligent energy management controller is built, which consists of two neural network (NN) modules that are trained based on the optimized results obtained by DP methods, considering the trip length and duration.

Numerical simulation shows that the proposed controller can improve the fuel economy of the vehicle. In Ref. [10], to achieve the optimal energy allocation for the engine-generator, battery and ultracapacitor of a plug-in hybrid electric vehicle, a novel adaptive energy management strategy has been proposed. This approach uses a fuzzy logic controller to classify typical driving cycles into different driving patterns and to identify the real-time driving pattern. Dynamic programming has been employed to develop optimal control strategies for different driving blocks, and it is helpful for realizing the adaptive energy management for real-time driving cycles. The simulation results indicate that the proposed energy management strategy has better fuel efficiency than the original and conventional dynamic programming-based control strategies.

At the same time, multi-objective optimal control has been widely used in the field of vehicles.In Ref. [11], a novel control strategy is proposed to manage the power distribution between the battery and UC for a hybrid energy management system (HEMS) in PEVs. This control strategy aims at realizing less power loss, longer battery lifecycle, as well as UC's stable terminal voltage and ability of quick charge/discharge. Based on these three optimization targets, the authors define three sets of loss functions and formulate a multi-objective optimization (MOO) problem to solve the problem. In Ref. [12], to improve the total efficiency of the drive system and the driving safety of distributed electric drive vehicles, this paper proposes a multi-objective optimization method based on torque allocation optimization. The linear weighting method with adaptive weight coefficients is used to transform the solution of the above two objective functions into a multi-objective optimization problem under constraint conditions. The authors solve the multi-objective optimization problem to obtain the optimal torque distribution of the distributed electric drive system. In Ref. [13], it provides an optimal selection methodology for plug-in hybrid electric vehicle (PHEV) powertrain configuration by means of optimization and comprehensive evaluation of powertrain design schemes. To determine performance potential, a configuration-sizing-control strategy integrated multi-objective powertrain optimization design is proposed and applied to series, parallel pre-transmission (P2), output power-split, and multi-mode power-split powertrain configurations. Considering simultaneous optimization of fuel economy, electric energy consumption, and acceleration capacity, the parameters of the powertrain components and vehicle performance of each configuration are optimized based on global optimal control in different situations of object trade-off. The results suggest that the P2 configuration and its optimal sizing can be selected when the goal is to optimize acceleration capacity, the multi-mode power-split configuration and its optimal sizing can be selected when the goal is to optimize electric energy efficiency, and the output power-split configuration and its optimal sizing can be selected when the fuel economy needs to be optimized.

There are also many researches combining multi-objective optimization and other methods of control strategies to solve different problems. In Ref. [14], a new particle swarm-based optimization method is presented for multi-objective optimization of vehicle crashworthiness. In Ref. [15], it presents a novel vehicular adaptive cruise control (ACC) system that can comprehensively address issues of tracking capability, fuel economy and driver desired response. Multi-objective optimization is synthesized under the framework of model predictive control (MPC) theory. In Ref. [16], a hybrid predictive control formulation based on evolutionary multi-objective optimization to optimize real-time operations of public transport systems is presented. In this case, the optimization was defined in terms of two objectives: waiting time minimization on one side, and impact of the strategies on the other. A genetic algorithm method is proposed to solve the multi-objective dynamic problem. In Ref. [17], an adaptive cruise control algorithm with multi-objectives is proposed based on a model predictive control (MPC) framework. In the proposed ACC algorithm, safety is guaranteed by constraining the inter-distance within a safe range; the requirements of comfort and car-following are considered to be the performance criteria and some optimal reference trajectories are introduced to increase fuel-economy.

## 2. The electro-mechanical transmission system

The electro-mechanical transmission system is composed of two components called mechanical transmission and electric power transmission, in which the power flow of the engine is separated and converged by the power coupling mechanism, as shown in figure 1. According to different energy forms of power transfer, EMT can work at mechanical transmission mode, electric power transmission mode and electro-mechanical transmission mode.

When working at electro-mechanical transmission mode, the engine power is divided to two parts: one of which is transmitted in the form of mechanical power, and the other is converted into electric power by generator. The electric power is also divided into several parts, one part is stored in the storage device, another is modulated into mechanical energy by motor, and the else is consumed by electric devices. After all, the directly transmitted mechanical power and motor power is converged to the driving wheel. The power flow keeps balance during separation, convergence, transmission and conversion. Due to the relationship of system components and the law of conservation of energy, the static power balance in electro-mechanical transmission mode can be expressed as follows:

The electric power balance equation:

$$P_S + P_c + P_l + P_A \eta_A^{-\text{sgn}(P_A)} + P_B \eta_B^{-\text{sgn}(P_B)} = 0 \qquad (1)$$

The driving power balance equation:

$$P_e \eta_e = P_d + P_c + P' + P_S \qquad (2)$$

Where $P_c$ is the electric power demand which is got by the electric condition. $P_l$ is the electric power loss in the process of transmission, $\eta_A$、$\eta_B$ are the power conversion efficiency of motor A and motor B, which can be obtained through the look-up table. $P_S$、$P_A$、$P_B$、$P_e$ are respectively the power of storage device, motor A, motor B and the engine. The battery charging power is assumed to be positive. $P_d$ is the demanded driving power. $\eta_e$ is the engine efficiency. $P'$ is the power loss of the whole process.

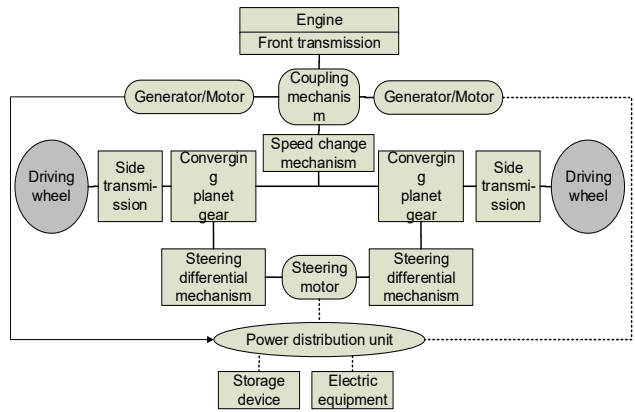

**Figure 1** Schematic diagram of the EMT structure

## 3. Driving condition division and selection of optimization objectives

### 3.1 Vehicle operating requirements

Mobility is the capability of the special type vehicle to move under certain conditions. Mobility can be divided into two categories: campaign mobility and tactical mobility. Campaign mobility is mainly indicated by on-road speed and maximum range. Tactical mobility, namely the capability of vehicles to move fast under all kinds of ground and terrain conditions that might be met under certain tactical background, is mainly indicated by acceleration, steering, braking and passability performance. In addition, tracked vehicles usually drive on harsh operating conditions. In order to meet operation requirements of special purpose, vehicles must have good passability and adaptability in off-road operating conditions.

2) Electric power generation capacity. On-board high-power electrical equipment and electric weapons is the future trend of development of tracked vehicles. EMT system should have adequate generation capacity to meet the real-time electric power demand, such as powering engine cooling fan, active suspension driving motor, and electric weapons, etc. Except for the current power demand, the EMT system should have some reserve electric power available at any time to meet the needs of the active suspension and electric weapons.

3) Power battery pack service life. Heavy vehicles have a relatively high demand for power level of power battery pack. However, with current technologies, the cost of power battery pack is high. Therefore, the power battery service life should be improved as much as possible. The SOC should remain basically unchanged.

4) Fuel economy. Decreasing the fuel consumption has not only economic significance, but also has military significance. As a type of electro-mechanical transmission, EMT system should decrease the fuel consumption as much as possible on condition that other properties are already met.

5) Other properties. Other properties such as the transmission efficiency of the transmission system, braking energy recovery performance, etc., are all required by EMT vehicles.

## 3.2 The selection of optimization objectives

Based on analysis of operating requirements of tracked vehicle above, it is concluded that the control strategy of power allocation is a multi-objective optimization problem. In order to obtain the best overall performance, many aspects of performance and vehicle characteristics must be considered during the process of building diversified optimization objectives.

1) Fuel economy. EMT system should take into account the usage of both engine and battery synthetically, thereby improve efficiency, fuel economy and service life. Therefore, $J_1$, the comprehensive economic optimization objective, is established as a function of power battery life service and engine fuel consumption:

$$J_1 = fuel + \gamma_1 \Delta SOC + \gamma_2 (SOC - SOC_0)^2 \quad (3)$$

Where fuel is the fuel consumption rate of the engine and the unit is g/s. fuel is a function of the engine speed and torque. $\gamma_1 \Delta SOC$ is the battery equivalent fuel consumption. $\gamma_1$ is the equivalent coefficient whose value is -12500. $\Delta SOC$ is the change of the state of charge. $\gamma_2(SOC-SOC_0)^2$ is the loss of power battery pack service life. $\gamma_2$ is a scaling factor. The bigger it is, the more sensitive $\gamma_2(SOC-SOC_0)^2$ is to $SOC$ changes. Its value taken here is 2000. $SOC$ is the state of charge of the power battery pack. $SOC_0$ is the initial state of charge and in this paper it is 0.5.

2) Power performance. Vehicle power performance can be measured by top speed, acceleration ability and grade ability. Since heavy tracked vehicles confront a wide variation range of load and grade, these two parameters affect a lot on mobility of heavy tracked vehicles. When the vehicle load or the road slope increases, more power is needed to ensure normal operation of the vehicle. Grade ability and acceleration ability are embodiments of power. They can be measure by reserve driving power.

$$J_2 = P_{emax}(n_e) + P_{Smax}(SOC) - P_d \quad (4)$$

Where $P_d$ is the driving power demand of the EMT system and the unit is kW. $P_{emax}$ is the maximum power the engine can provide and the unit is kW. $P_{emax}$ is a function of engine speed $n_e$. The bigger the $P_{emax}$ is, the bigger the $J_2$ is and the better the power performance is. $P_{Smax}$ is the maximum battery discharge power and the unit is kW. $P_{Smax}$ is a function of battery SOC.

3) Power generation capacity. This property is similar to dynamic performance. On condition that the EMT system meets current power requirement, $J_3$ is the difference between the maximum power EMT can provide currently and the power needed by the EMT system.

$$J_3 = P_{Amax}(T_A) + P_{Bmax}(T_B) + P_{Smax}(SOC) - P_c \quad (5)$$

Where $P_c$ is the power requirement of the EMT system and the unit is kW. $P_{Amax}$ is the maximum power generating of generator A and the unit is kW. $P_{Amax}$ is a function of generator torque $T_A$. $P_{Bmax}$ is the maximum generating power of generator B and the unit is kW. $P_{Bmax}$ is a function of generator torque $T_B$. $P_{Smax}$ is the maximum battery discharge power and the unit is kW. $P_{Smax}$ is a function of battery SOC.

At present, the essence of dealing with multi-objective optimization problem is to transfer each sub-objective function into a single-objective function or other simpler multi-objective function by processing or mathematical transformation. The main methods are evaluation function method, hierarchical method and interactive programming method. In this paper, multi-objective problems will be transferred into comprehensive single-objective problems by adopting weighted method. Under different operating conditions, different coefficients are selected to meet different economic, dynamic and generating requirements.

## 3.3 Operating condition division

For passenger cars, operating conditions can be divided into three categories: city, suburb and highway. For tracked vehicles, operating conditions are more complicated and changeable. There are potholes, soft road, flat road, gradient road, narrow road and wading road, etc. Currently there is no standard clearly definition on operating conditions of tracked vehicles. In this paper, tracked vehicle operating condition will be divided according to vehicle operating environment and operating requirements, basing on experimental data.

1) Low speed driving condition. The driving environment of tracked vehicles is mostly off-road environment, where the road drag coefficient is big. In order to guarantee vehicle passability and safety, vehicles mostly run at low speed to ensure sufficient reserve power. At the same time, in order to ensure accurate shooting, electronic weapons are mostly used at low speed. On this condition, the military requirements and the vehicle passabilty are the main objectives. So the vehicle must ensure adequate power supply and driving capability. Based on experimental data and prior knowledge, the speed range of 0-35km/h is considered as low speed driving condition.

2) Medium speed driving condition. According to the operational requirements and strategic deployment, tracked vehicles often need rapid transfer of strategic position. At the same time, in order to raise driving range or radius of action, fuel consumption should be minimized under the premise of ensuring dynamic. On this condition, the engine should work in high efficiency area to reduce fuel consumption. In order to prolong the service life of the power battery pack, the SOC should remain unchanged as much as possible. At this time, the power of the engine outputs to driving wheels through power coupling mechanism mainly in the form of mechanical power, and thereby meets vehicle driving requirement on good road. Based on experimental data and prior knowledge, the speed range 35-60km/h is considered as medium speed driving condition.

3) High speed driving condition. Tracked vehicles can realize fast transfer on cement or asphalt pavement. On the battlefield, time is very precious. When executing evacuation or attacking tasks, vehicles should be able to move at the fastest speed to occupy the initiative. Under this condition, the power of the engine and the battery pack output together to driving wheels through power coupling mechanism, and thereby meets the high power demand. Driving speed higher than 60km/h is considered as high speed driving condition.

## 4. The comprehensive optimization objective based on analytic hierarchy process

The optimization objectives are changed into a comprehensive optimization objective based on the method of weighting, in which the weight coefficients are the key of multi-objective problems. According to different demands of vehicle driving conditions, specific optimization objectives are confirmed by using AHP.

### 4.1 Analytic hierarchy process(AHP)

The Analytic Hierarchy Process(AHP) is a process in which decision-makers make their decision process modelling and quantification[5,6]. It assists the decision maker to solve the problem by decomposing a complex problem into a multilevel hierarchic structure of objectives, criteria, subcriteria and alternatives and provides a fundamental scale of relative magnitudes expressed in dominance units to represent judgments in a convincing way.

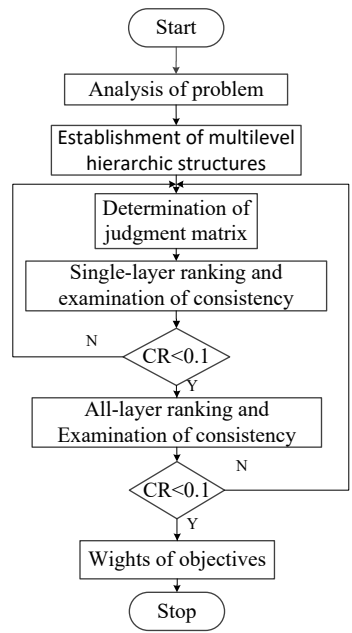

**Figure 2** Solving process of AHP

The fundamental of AHP is evaluating schemes according to the hierarchic structure of objectives, criteria, sub-criteria and constraint conditions, confirming judgment matrix in the form of pairwise comparisons, using eigenvector corresponding to the maximum eigenvalue as the weight coefficients. In this way, weights(priority) of all schemes can be determined, as shown in figure 2:

## 4.2 The comprehensive multi-objectives based on AHP

### 4.2.1 Establishment of multilevel hierarchic structures

Generally speaking, elements of upper layer dominate all or some of the elements of the adjacent lower layer, thus forming the dominance relation from the upper layer to the next, while elements in the same layer do not have the dominance relation or dependence relationship. Layers with such characters are called hierarchic layers. There are three classes when grouping by problem factors and regarding each group as a layer, as shown in figure 3.

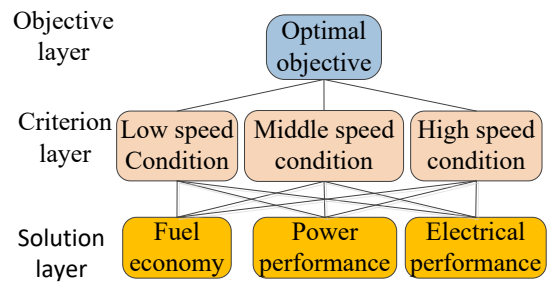

**Figure3** multilevel hierarchic structures

The top layer is the layer of comprehensive optimization objectives which has a single element. It is called objective layer since it is the intended objective or optimal result of the problem. The middle layer is the criterion of working conditions, it includes intermediate process to accomplish the targets and consists of a number of layers such as criteria and sub-criteria. Therefore, the middle layer is also called the criterion layer. The bottom layer consists of sub-objectives of optimization including alternative measures and solutions to accomplish the targets, so it is called the solution layer.

## 4.2.2 Establishment of the judgment matrix

The judgment matrix is used to evaluate the relative importance of the elements in the same layer. To confirm the impact of n elements $y=(y_1, y_2,…, y_n)$ on the objective $z$, weights of elements are obtained by comparison matrix which is made up of pairwise relative importance. In the comparison matrix $A=(a_{ij})n×n$, $a_{ij}$ is the relative importance of element $y_i$ and $y_j$ on effects to $z$.

$$A = \begin{pmatrix} a_{11} & a_{12} & \cdots & a_{1n} \\ a_{21} & a_{22} & \cdots & a_{2n} \\ \vdots & \vdots & & \vdots \\ a_{n1} & a_{n2} & \cdots & a_{nn} \end{pmatrix} \quad (6)$$

Where $a_{ij}>0$, $a_{ji}=1/a_{ij}$, $a_{ii}=1$, $(i, j=1,2,…,n)$. On such assumption, A is called positive reciprocal matrices. Saty provides a methodology to quantify by the scale range from number 1 to 9 and their reciprocals, as shown in figure 1.

Under low speed driving conditions, with the use of weapons such as electric weapons and active suspension, vehicle power demand is large. At the same time, power performance is required to satisfy the trafficability characteristic and motility under complex pavement conditions. Battle requirements become a priority under such conditions, power generation performance and motility should have priority over fuel economy. Based on the scale as shown in table 1, $a_{21}=5, a_{31}=9, a_{32}=7$ are confirmed and there is the coefficient matrix under such conditions.

$$A_1 = \begin{pmatrix} 1 & 1/5 & 1/9 \\ 5 & 1 & 1/7 \\ 9 & 7 & 1 \end{pmatrix}$$

Similarly, according to the vehicle driving demand under different conditions as mentioned, corresponding scales are chosen and coefficient matrices under medium and high speed conditions are confirmed.

$$A_2 = \begin{pmatrix} 1 & 5 & 9 \\ 1/5 & 1 & 3 \\ 1/9 & 1/3 & 1 \end{pmatrix}, \quad A_3 = \begin{pmatrix} 1 & 1/9 & 1/3 \\ 9 & 1 & 7 \\ 3 & 1/7 & 1 \end{pmatrix}$$

**Table1** Scales

| Scale | Meaning |
|---|---|
| 1 | Comparing two factors, the two factors have the same importance. |
| 3 | Comparing two factors, the former factor is slightly more important than the latter. |
| 5 | Comparing two factors, the former factor is obviously more important than the latter. |
| 7 | Comparing two factors, the former factor is strongly more important than the latter. |
| 9 | Comparing two factors, the former factor is extremely more important than the latter. |
| 2/4/6/8 | Median of adjacent judgment mentioned above(Less application) |

### 4.2.3 layer ranking

1) The single layer ranking is the basis of the ranking of the impact that all the elements of the layer have on the upper layer. It is intended for confirming the impact ranking of this layer's elements related to a certain element of the upper layer.

Set the weight vector as $\boldsymbol{\theta} = [\theta_1, \theta_2, ..., \theta_n]^T$, then:

$$\mathbf{A}\boldsymbol{\theta} = \lambda\boldsymbol{\theta} \quad (7)$$

Where λ is the maximum positive eigenvalue of A, and θ is the eigenvector of A corresponding to λ. Thereby the single layer ranking can be turned into a computation of eigenvalue $\lambda$ and corresponding eigenvector of the judgment matrix, and the relative weights of the scales in this layer are confirmed.

2) The total layers ranking. The impact weights of all the elements in this layer to the upper can be calculated by using the result of all the single layer rankings in this layer. The total layers ranking must be calculated from the upper layer to the next, after finishing the total layers ranking of the upper layer $A_1$, $A_2$,.., $A_m$, the weights of elements are got as [$a_1$,$a_2$,...,$a_m$]. Set the single layer ranking of elements $B_1$,$B_2$,..,$B_n$ corresponding to $a_j$ as $[b_1^j, b_2^j, ..., b_n^j]^T$, then $b_i^j = 0$ if $B_i$ has no association with $A_i$. Then the total layers ranking corresponding to $B_i$ is $b_i = \sum a_j b_i^j$.

3) Computation of eigenvalue and corresponding eigenvector. There are convenient methods to calculating the eigenvalue of the judgment matrix $A_1$, such as(和法、方根法和幂法). In this paper, 和法 is used to calculate the eigenvalue and corresponding eigenvector of the judgment matrix $A_1$, as described below:

 a. Normalize each column of matrix $\mathbf{A}_1$: $\tilde{\theta}_{ij} = a_{ij} / \sum_{i=1}^{n} a_{ij}$;

 b. Sum $\tilde{\theta}_{ij}$ by row: $\tilde{\theta}_i = \sum_{j=1}^{n} \tilde{\theta}_{ij}$;

 c. Normalization: $\tilde{\theta} = (\tilde{\theta}_1, \tilde{\theta}_2, ..., \tilde{\theta}_n)^T$ and the eigenvector is: $\theta = (\theta_1, \theta_2, ...\theta_n)^T, \theta_i = \tilde{\theta}_i / \sum_{i=1}^{n} \tilde{\theta}_i$;

 d. Calculate $\mathbf{A}_1\theta$;

 e. The approximate value of the biggest eigenvalue $\lambda$ can be calculated as $\lambda = \frac{1}{n}\sum_{i=1}^{n} \frac{(\mathbf{A}_1\theta)_i}{\theta_i}$

After calculating by the method mentioned above, the biggest eigenvalue $\lambda_{max}$ of matrix $A_1$ is 3.08 and the corresponding eigenvector is $(0.05, 0.66, 0.29)^T$; the biggest eigenvalue $\lambda_{max}$ of matrix $A_2$ is 3.03 and the corresponding eigenvector is $(0.67, 0.27, 0.06)^T$; the biggest eigenvalue $\lambda_{max}$ of matrix $A_3$ is 3.08 and the corresponding eigenvector is $(0.15, 0.78, 0.07)^T$.

## 4.2.4 Examination of consistency

Although the judgment matrix in the form of pairwise comparisons do not always maintain consistency, it is necessary to calculate the consistency index *CI* for consistency examination. When *CI*=0, the matrix is completely consistent matrix, otherwise, the bigger the *CI* is, the worse the consistency of the judgment is.

$$CI = \frac{\lambda_{max} - n}{n - 1} \quad (8)$$

To examine the consistency of the judgment matrix, the comparison between *CI* and the mean random consistency index *RI*(as shown in table 2) is needed. The result of comparing the matrix consistency index *CI* with the same order mean random consistency index *RI* is called random consistency ratio, written as *CR*. Generally, when

$$CR = \frac{CI}{RI} < 0.1 \quad (9)$$

the judgment matrix satisfying (9) are considered to have qualified consistency, otherwise, adjustment needs to be taken until it satisfied.

**Table2** The mean random consistency index *RI*

| Order | 1 | 2 | 3 | 4 | 5 | 6 | 7 | 8 |
|---|---|---|---|---|---|---|---|---|
| RI | 0 | 0 | 0.58 | 0.9 | 1.12 | 1.24 | 1.32 | 1.41 |
| Order | 9 | 10 | 11 | 12 | 13 | 14 | 15 | |
| RI | 1.45 | 1.49 | 1.52 | 1.54 | 1.56 | 1.58 | 1.59 | |

The consistency index and random consistency ratio of coefficient matrix $A_1$ are as follows:

$$\begin{cases} CI = \frac{\lambda_{max} - n}{n - 1} = \frac{3.08 - 3}{3 - 1} = 0.04 \\ CR = \frac{CI}{RI} = \frac{0.04}{0.58} = 0.069 \end{cases} \quad (10)$$

Similarly, the consistency index of $A_2$ and $A_3$ are 0.026 and 0.069, both of which are smaller than 0.1. Thus the consistency of judgment matrices $A_1$, $A_2$ and $A_3$ are confirmed to have satisfying consistency.

## 4.3 Normalization

The optimization objectives calculated by formulas (3) - (5) have the differences of units and orders of magnitudes so that the weighting method cannot be directly used, and because of this, normalization becomes necessary[7]. After normalization, every objective is in the same order of magnitude and the sum of weights equals 1. In this paper, every optimization objective is unified by normalization, as described below:

1) Equivalent fuel economy $\bar{J}_1$. According to the engine fuel consumption rate, the maximum fuel consumption rate is fuelmax=72g/s. The SOC of power battery is:

$$\Delta SOC = \frac{\sqrt{V_{oc}^2 + 4P_S R_b} - V_{oc}}{7200 C_b R_b} \quad (11)$$

According to the maximum charging(discharging) power of the power battery $P_{Smax}$=220kW, the maximum variation range is $\Delta SOC_{max}$=0.001. According to the variation range [0.3,0.8] and the initial value $SOC_0$=0.5, the maximum ($SOC$-$SOC_0$) is 0.3. equivalent fuel economy $\bar{J}_1$ thereby is defined as:

$$\bar{J}_1 = \frac{J_1}{J_{1max}} = \frac{fuel + \gamma_1 \Delta SOC + \gamma_2 (SOC - SOC_0)^2}{fuel_{max} + \gamma_1 \Delta SOC_{max} + \gamma_2 (SOC - SOC_0)^2_{max}} \quad (12)$$

2) Equivalent maneuverability $\bar{J}_2$. Equivalent maneuverability is the reciprocal of the ratio of J2 in current state and the maximum driving power system can provide, that is:

$$\bar{J}_2 = -\frac{J_2}{J_{2max}} = -\frac{P_{emax}(n_e) + P_{Smax}(SOC) - P_d}{P_{emax}(n_e) + P_{Smax}(SOC)} \quad (13)$$

Where $P_{emax}$、$P_{Smax}$ are determined by state quantities, not constant values.

3) Equivalent generating capacity $\bar{J}_3$. Similarly with the calculation of equivalent maneuverability, $\bar{J}_3$ is the reciprocal of the ratio of $J_3$ in current state and the maximum electric power demand, that is:

$$\bar{J}_3 = -\frac{J_3}{J_{3\max}}$$
$$= -\frac{P_{A\max}(T_A) + P_{B\max}(T_B) + P_{S\max}(SOC) - P_c}{P_{A\max}(T_A) + P_{B\max}(T_B) + P_{S\max}(SOC)} \quad (14)$$

Where $P_{A\max}$、$P_{B\max}$、$P_{S\max}$ are determined by state quantities, not constant values.

After normalization of each optimization objective, weights under different conditions can be calculated by using AHP, so the comprehensive optimization objective under different conditions is:

$$\begin{cases} V_1 = 0.05\bar{J}_1 + 0.29\bar{J}_2 + 0.66\bar{J}_3 \\ V_2 = 0.67\bar{J}_1 + 0.27\bar{J}_2 + 0.06\bar{J}_3 \\ V_3 = 0.15\bar{J}_1 + 0.78\bar{J}_2 + 0.07\bar{J}_3 \end{cases} \quad (15)$$

## 5. Control strategy based on dynamic programming

### 5.1 Dynamic programming algorithm

In recent years, dynamic programming (DP) is widely used in the field of hybrid energy management. The main idea is to transfer complex optimal control problems into recursive functions of multistage decision process and the object must has unfollow-up effect property. The basis and the core of DP is the principle of optimality whose contents are the rest decisions are optimal with respect to the states resulted from the initial decisions, no matter what the initial stat and initial decisions are.

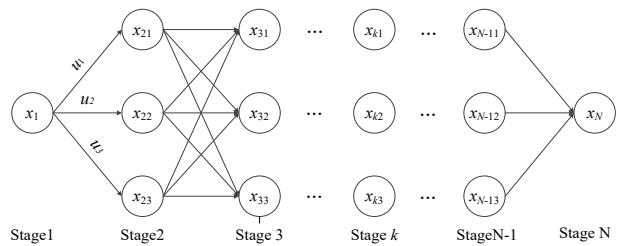

**Figure 4** Diagrammatic sketch of DP

The essence of DP is division and eliminate redundancy, that is to say, divide the problem into small and similar sub-problems and save the solutions of sub-problems to avoid repetitive computation, thus the optimal problem can be solved. Sub-optimal schemes are excluded in the process by establishing the Bellman equation and adopting backward solving method. As a result, the computation workload is deduced greatly compared with the method of exhaustion.

## 5.2 Control model based on DP

The driving data of tracked vehicle under realistic road conditions have been researched, as shown in figure 5, which includes vehicle speed $v$ and road dragging coefficient f. There are 5 different driving conditions, the corresponding electric power demand of which are as shown in figure 6.

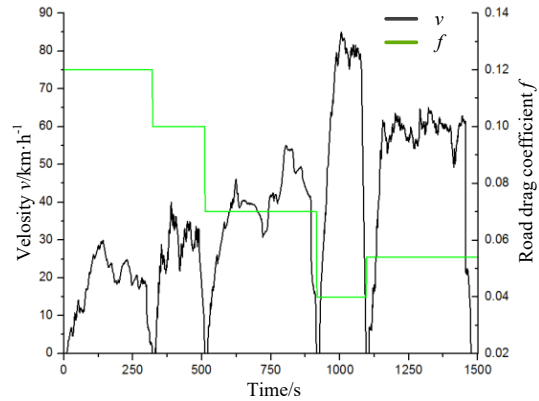

**Figure 5** vehicle drive cycle

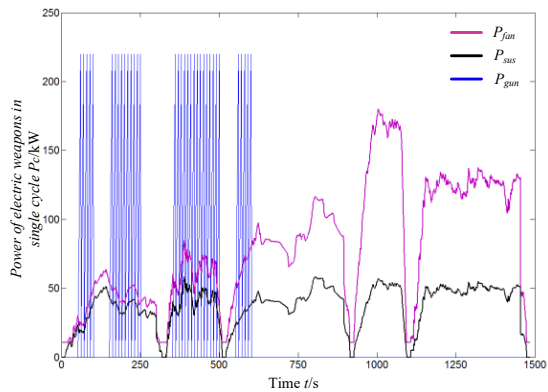

**Figure 6** Power demand curve

The condition is divided into N=1486 states by time scale. The engine torque, speed of motor A and B, the vehicle speed and the power battery are chosen as the state variables, that is x=($T_e$, $n_A$, $n_B$, $SOC$). The engine speed, torque of motor A and torque of motor B are the decision variables, that is a=($n_e$, $T_A$, $T_B$). The optimization objective is a comprehensive performance function $V = \alpha_1 \bar{J}_1 + \alpha_2 \bar{J}_2 + \alpha_3 \bar{J}_3$, where $\alpha_1, \alpha_2, \alpha_3$ are the weights of fuel economy, dynamic Performance and generating capacity.

Set $r(x_k, a_k)$ as single-step cost function while making decision $a_k$ in stage k with the state $x_k$:

$$r(x_k, a_k) = \alpha_1 \bar{J}_1(x_k, a_k) + \alpha_2 \bar{J}_2(x_k, a_k) + \alpha_3 \bar{J}_3(x_k, a_k) \tag{16}$$

Then the cost function of the whole process from stage *k* to stage *N*:

$$\begin{aligned} V(x_k) &= r(x_k, a_k) + r(x_{k+1}, a_{k+1}) + ... + r(x_{N-1}, a_{N-1}) \\ &= \sum_{i=k}^{N-1} r(x_i, a_i) \end{aligned} \tag{17}$$

This is an optimal control problem, in other words, an optimal controlling vector A*(k) should be found in the available control set A to make the cost function achieve minimum value $V^*(x_k)$ from stage $x_k$ to the final stage $x_N$ in the time scale[$t_k, t_N$]. According to the principle of optimization, the recursion equation of the optimal cost function, namely the Bellman equation, are deduced as:

$$\begin{aligned} V^*(x_k) &= \min_{\pi=\{a_k, a_{k+1},...,a_{N-2}, a_{N-1}\}} \left\{ r(x_k, a_k) + \sum_{i=k+1}^{N-1} r(x_i, a_i) \right\} \\ &= \min_{a_k \in \pi} \min_{\pi_{k+1}=\{a_{k+1},...,a_{N-2}, a_{N-1}\}} \left\{ r(x_k, a_k) + \sum_{i=k+1}^{N-1} r(x_i, a_i) \right\} \\ &= \min_{a_k \in \pi} \{ r(x_k, a_k) + V^*(x_{k+1}) \} \end{aligned} \tag{18}$$

The multistage decision problem is divided into some sub-problems, and the global optimal control strategy A*(*k*) can be got by using reverse iterative method. The basic procedure solving the optimal problem by using DP is shown in figure 7.

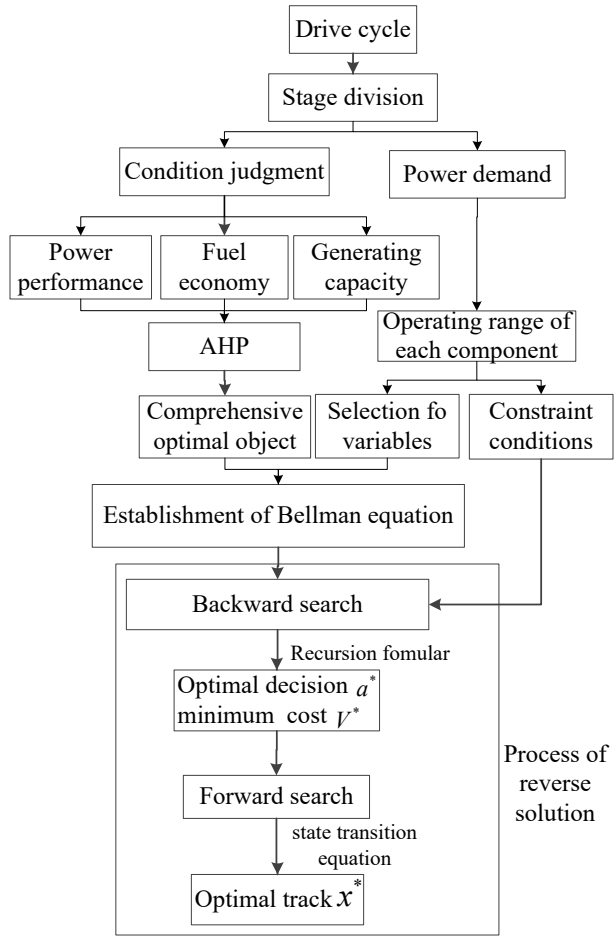

**Figure 7** Multi-objective control strategy based on driving pattern division

## 6. Simulation and result analysis

### 6.1 EMT simulation model

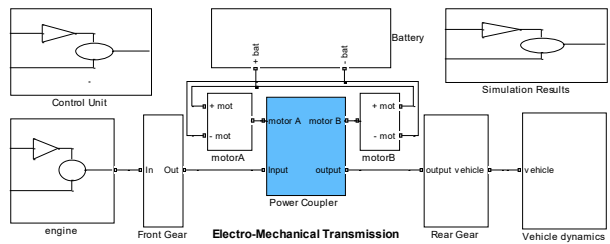

**Figure 8** Simulation of EMT system

EMT system simulation model is built in Matlab / Simulink environment. The model mainly consists of the engine, the power coupling mechanism, the motor, etc. In addition, there are controller subsystem and simulation result display subsystem as well. The model of each subsystem is built by SimpowerSystem and Simdriveline module library. These subsystems are connected by a mechanical axis. Figure 8 shows an example of EMT system simulation model.

## 6.2   Analysis of the simulation result

Comparison has been made between the control strategy presented in this paper and the strategy based on rules which is used before on the research platform. The power changing curves of each component in both control strategy are shown in figure 9. According to figure 9, the DP control strategy based on driving pattern division tends to satisfy the vehicle power demand with engine power, and the power battery is used to compensate for the dynamic characteristics of engine and realize braking energy recovery. And the two motors, whose power difference value supplies electrical equipments and maintains the state of power battery, are used to help the engine adjust the working point. As a result, the power battery has sufficient capacity to discharge and get charged in the long time. On the other hand, it is beneficial for the power battery's service life to be maintained at a point where the system keeps the balance of charge and discharge.

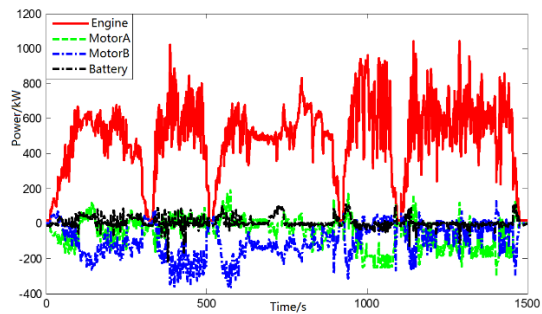

(a) Power changing curve under multi-objective control strategy

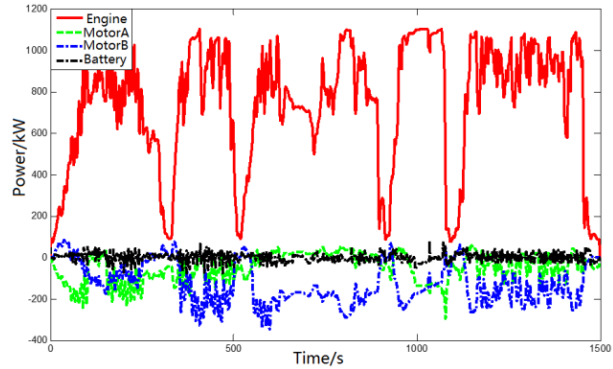

(b) Power changing curve under rule-based control strategy

**Figure 9** Comparison of power changing curve

The distributions of engine working points in different control strategy are shown in figure 10. The vehicle drive cycle includes three conditions which focus on different objectives. On low speed conditions the focuses are power performance and generating capacity; on middle speed conditions the focus is fuel consumption, while on high speed conditions the focus is vehicle maneuverability. Vehicle drive cycles are mainly focused on middle to low speed range and rarely distribute on high speed range. As shown in figure 10(a), to have a good power performance and fuel consumption in low speed conditions, the DP control strategy increases the engine power. As a result, there are two trends. One is the working points of the engine in low speed range are higher than the most economical fuel consumption curve and near to the external characteristic curve, and the other one is to fit to the maximum power speed which is in middle or high speed range. By this way, the fuel economic performance can be improved as much as possible on condition that the power performance and the generating capacity are satisfied. According to figure 10(b), the working points under rule-based control strategy focus on the speed range [2700，4200] and are close to the engine external characteristic curve so that the power performance and generating capacity of the engine can be guaranteed. The fuel consumption of engine under DP multi-objective control strategy is 68.6906 L, and 81.3792L under rule-based control strategy; the fuel consumption performance is improved by 16%.

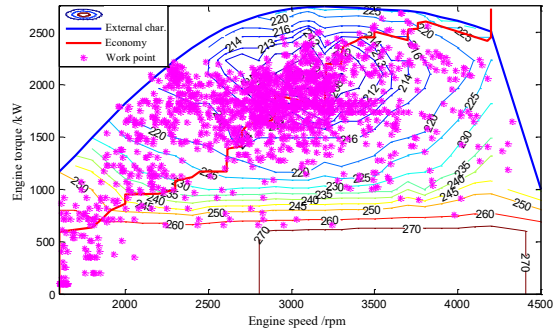

(a) Operating points of engine under multi-objective control strtegy

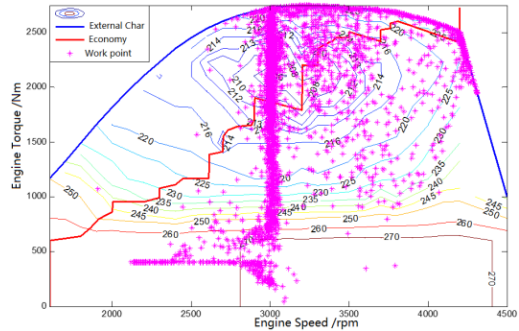

(b) Operating points of engine under rule-based control strategy

Figure 10 Distribution of operating points of engine

## 7. Conclusion

1) This paper analyzes the operating requirements of tracked vehicles and divides driving conditions of tracked vehicles. The paper also explains how to use AHP to get optimization objective weight coefficients on various operating conditions and then use weighted sum method to transform multiple optimization objectives into one single comprehensive optimization objective. The paper proposes a new solution for multi-objective optimization problems.

2) Using the DP multi-objective control strategy proposed in this paper , which is based on operating condition division, the global optimal control strategy can be obtained. Through simulation and comparison, it is verified that after adopting this strategy, the fuel economy and the comprehensive performance are both improved obviously.

## Acknowledgement

This work was supported by the National Natural Science Foundation of China. (51005017, 51575043 and U1564210)